\documentclass[conference]{IEEEtran}
\IEEEoverridecommandlockouts
\usepackage{booktabs} 
\usepackage{soul}
\usepackage{threeparttable}
\usepackage{enumitem}
\usepackage{indentfirst}
\usepackage{multirow}
\usepackage{makecell}
\usepackage{balance}
\usepackage{cite}
\usepackage{caption}
\usepackage{amsmath,amssymb,amsfonts}
\usepackage{algorithm}
\usepackage{algpseudocode}
\usepackage{tikz}
\usepackage{newunicodechar}

\newcommand{\blackcircled}[1]{%
  \tikz[baseline=(char.base)]{
    \node[shape=circle, fill=black, text=white, inner sep=0.4pt] (char) {\sffamily\bfseries\small #1};
  }%
}

\newunicodechar{❶}{\blackcircled{1}}
\newunicodechar{❷}{\blackcircled{2}}
\newunicodechar{❸}{\blackcircled{3}}
\newunicodechar{❹}{\blackcircled{4}}

\usepackage{graphicx}
\usepackage{textcomp}
\usepackage{xcolor}
\usepackage{booktabs}
\usepackage{array}      
\usepackage{makecell}   

\def\BibTeX{{\rm B\kern-.05em{\sc i\kern-.025em b}\kern-.08em
    T\kern-.1667em\lower.7ex\hbox{E}\kern-.125emX}}
\begin{document}

\title{
NEURAL: An Elastic \underline{Neur}omorphic \underline{A}rchitecture with Hybrid Data-Event Execution and On-the-f\underline{l}y Attention Dataflow
}
\author{
    Yuehai Chen, and Farhad Merchant\\
    Bernoulli Institute and CogniGron, University of Groningen, The Netherlands\\
    Email: \{yuehai.chen, f.a.merchant\}@rug.nl
}
\maketitle

\begin{abstract}

Spiking neural networks (SNNs) have emerged as a promising alternative to artificial neural networks (ANNs), offering improved energy efficiency by leveraging sparse and event-driven computation. However, existing hardware implementations of SNNs still suffer from the inherent spike sparsity and multi-timestep execution, which significantly increase latency and reduce energy efficiency.
This study presents NEURAL, a novel neuromorphic architecture based on a hybrid data-event execution paradigm by decoupling sparsity-aware processing from neuron computation and using elastic first-in-first-out (FIFO). NEURAL supports on-the-fly execution of spiking QKFormer by embedding its operations within the baseline computing flow without requiring dedicated hardware units. It also integrates a novel window-to-time-to-first-spike (W2TTFS) mechanism to replace average pooling and enable full-spike execution. Furthermore, we introduce a knowledge distillation (KD)-based training framework to construct single-timestep SNN models with competitive accuracy. 
NEURAL is implemented on a Xilinx Virtex-7 FPGA and evaluated using ResNet-11, QKFResNet-11, and VGG-11. Experimental results demonstrate that, at the algorithm level, the VGG-11 model trained with KD improves accuracy by 3.20\% on CIFAR-10 and 5.13\% on CIFAR-100. At the architecture level, compared to existing SNN accelerators, NEURAL achieves a 50\% reduction in resource utilization and a 1.97× improvement in energy efficiency. 

\end{abstract}


\section{Introduction}
Recently, an increasing number of edge devices have gained the capability to perform intelligent processing. Although current mainstream artificial neural networks (ANNs) demonstrate excellent performance, their deployment on resource-constrained edge devices remains challenging due to their complex computation and high power consumption. In contrast, spiking neural networks (SNNs) deliver information in binary spikes, which inherently exhibit low power consumption and event-driven computation. However, SNNs still have challenges such as high latency due to multi-timestep inference, binary signals that limit backpropagation, and general-purpose hardware struggling to efficiently support event-driven mechanisms. To address these issues, the algorithm and hardware co-design becomes the key to improving the execution efficiency of SNNs. 
\begin{figure}[htbp] 
\centering
\includegraphics[width=\columnwidth]{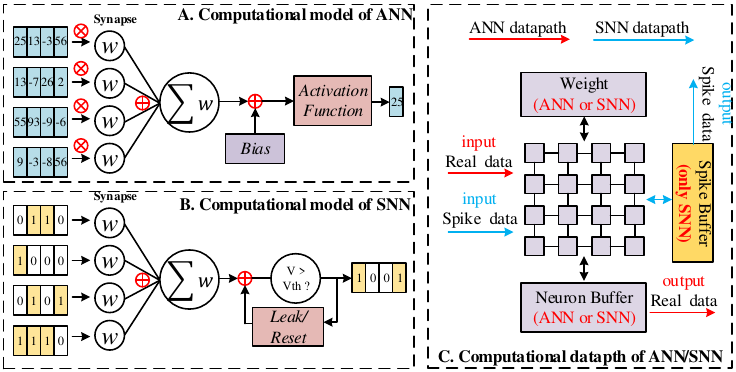} 
\caption{Differences between ANN and SNN computational models.}
\vspace{-20pt}
\label{fig}
\end{figure}
Fig. 1 shows the differences between the hardware implementation of ANNs and SNNs. Compared to ANNs, SNNs eliminate the need for multipliers and complex activation functions, relying instead on addition and comparator, but require additional memory to store membrane potentials and spike data for temporal processing. Researchers have proposed 3D architectures \cite{fang_energy-efficient_2025} \cite{chen_sibrain_2024} to process multiple time steps in parallel at the cost of high resource overhead, while sparsity-aware architectures \cite{chen_cerebron_2022} optimize computation taking advantage of the sparsity of spikes. In terms of model optimization, surrogate gradient methods \cite{wu_spatio-temporal_2018} have significantly reduced the number of time steps in recent years while maintaining accuracy. Knowledge distillation (KD) \cite{liang_knowledge_2024}\cite{yu_efficient_2025} has also been widely used to train low time-step SNNs to guide student networks to learn better representations via ANN teacher networks, achieving a good balance between low latency and high accuracy. Meanwhile, spike-based transformer mechanisms \cite{yao_spike-driven_2023} \cite{zhou_qkformer_2024} further improve the recognition accuracy of SNN models. However, most existing approaches remain at the algorithmic level, constructing merely spiking-like computation graphs without true spiking execution. This highlights the importance of algorithm-hardware co-design, emphasizing that optimization should be performed not only at the algorithmic level, but also at the architectural level, in order to fully exploit the efficiency potential of neuromorphic computing.
To further explore efficient SNN hardware architectures, we focus on building a co-optimized architecture with both full spiking computation and sparsity-aware capabilities that can perform inference in a single time step. Achieving high-accuracy single-timestep SNNs at the algorithmic level eliminates the need for multi-timestep scheduling in hardware, thereby reducing both inference latency and control logic complexity. Furthermore, identifying operations that remain non-spiking and transforming them into spike-based counterparts can further enhance the energy efficiency of event-driven computations.
Based on these observations, in this paper, we propose NEURAL, a hybrid data-event execution neuromorphic computing architecture that supports elastic connectivity and on-the-fly attention dataflow. The main contributions are as follows.
\begin{enumerate}[leftmargin=*]
\item A training framework combining KD and fixed-point quantization, enabling single-timestep SNNs to achieve competitive accuracy compared to multi-timestep models
\item Window-to-time-to-first-spike (W2TTFS) mechanism for converting non-spiking average pooling into spike-based computation while maintaining accuracy
\item Hybrid data-event execution using elastic FIFO scheduling to enable data-driven control and event-driven neuron computation, while supporting on-the-fly spiking QKFormer \cite{zhou_qkformer_2024} without dedicated hardware units
\item FPGA implementation of the NEURAL architecture deploying three deep SNNs, VGG-11, ResNet-11, and QKFResNet-11, achieves accuracies of up to 93.46\% and 72.1\% on CIFAR-10 and CIFAR-100 based on the KD training, respectively. Under a single-timestep execution paradigm, NEURAL significantly outperforms the existing STI-SNN \cite{wang_sti-snn_2025} architecture, exhibiting nearly a 3.9× improvement in computing efficiency.
\end{enumerate} 
\textcolor{black}{\indent The rest of this paper is organized as follows. Section II summarizes the related neuromorphic algorithms and architectures. Section III introduces the W2TTFS mechanism and the training framework based on knowledge distillation. Section IV describes the detailed design of the NEURAL architecture. Section V evaluates and analyzes the performance of NEURAL. Finally, Section VI concludes this paper and discusses future directions.}

\section{Related Works}
Neuromorphic computing evolves in both algorithms and architectures. At the algorithm level, the development of spiking models and training methods has enabled SNNs to approach the performance of ANNs. At the architecture level, the inherent event-driven computation and sparsity of SNNs are leveraged to enhance energy efficiency. Therefore, we summarize related work from both aspects.\\
\noindent\textbf{Neuromorphic algorithms.} With the development of the theory of spiking computation, the network structure has been expanded from the initial simple spiking multi-layer perception (SMLP) to the deep spiking convolutional neural network (DSCNN). Training methods have gone through the transition from bio-inspired spike-timing-dependent plasticity \cite{diehl_unsupervised_2015} (STDP) and ANN-to-SNN \cite{rueckauer_conversion_2017} to surrogate gradient-based supervised training methods \cite{wu_spatio-temporal_2018}. In order to further utilize the expressive power of ANN models in deep learning, several studies have explored the use of KD to improve the training of SNNs. Recent work shows that a ResNet-19 SNN model trained with KD can achieve 96. 65\% and 81. 47\% accuracy in the CIFAR-10 and CIFAR-100 datasets, respectively, using only 2 time steps \cite{yu_efficient_2025}, significantly improving the practical value of the low-timestep SNNs.\\
\noindent\textbf{Neuromorphic architectures.} Compared with general-purpose processors, dedicated neuromorphic architectures are better suited to the event-driven and sparse computational model of SNN, thus improving energy efficiency and throughput. Existing research on SNN accelerators can be categorized into the following directions: one class of work focuses on the efficient implementation of spiking MLPs \cite{chu_neuromorphic_2022} \cite{nguyen_low-power_2022}; the other class designs sparsity-aware execution mechanisms to leverage the inherent activation sparsity. Sparsity-aware execution mechanisms \cite{chen_cerebron_2022} \cite{aliyev_pulse_2024} can dynamically skip zero-valued computation and reduce redundant accesses, thus significantly improving computational efficiency. In addition, to mitigate the latency and computational overhead caused by the multi-timestep nature of SNNs, various studies have explored time-step compression techniques and time-parallel computing architectures \cite{fang_energy-efficient_2025} \cite{chen_sibrain_2024} \cite{9773259}. However, these approaches often introduce more complex resource scheduling and higher on-chip overhead while saving latency, making them difficult to deploy in resource-constrained edge environments. NEURAL bridges the gap between algorithmic improvements and architectural support, thereby addressing a major limitation of previous SNN hardware. It enables high-accuracy, low-latency, fully spike-based execution through co-optimized training and elastic hardware design with a smaller area, while supporting emerging SNN modules such as QKFormer in on-the-fly execution. This makes NEURAL a practical and scalable solution for next-generation neuromorphic computing systems.
\section{Full-Spike SNN Model}
SNNs transmit information through spikes, offering inherent advantages in computational and energy efficiency. However, current mainstream models struggle to achieve a fully spike-based computational path, particularly due to the use of average pooling in the downsampling stages. Although average pooling improves training stability under sparse input, it introduces continuous values that break the consistency of spike-based execution and increase computational and energy overhead. To address this, we propose the W2TTFS mechanism, which converts spike windows into time-to-first-spike representations across multiple timesteps during inference. By preserving fully spiking inputs to the classifier, this mechanism balances classification accuracy and hardware efficiency.
\begin{figure}[htbp] 
\centering
\includegraphics[width=\columnwidth]{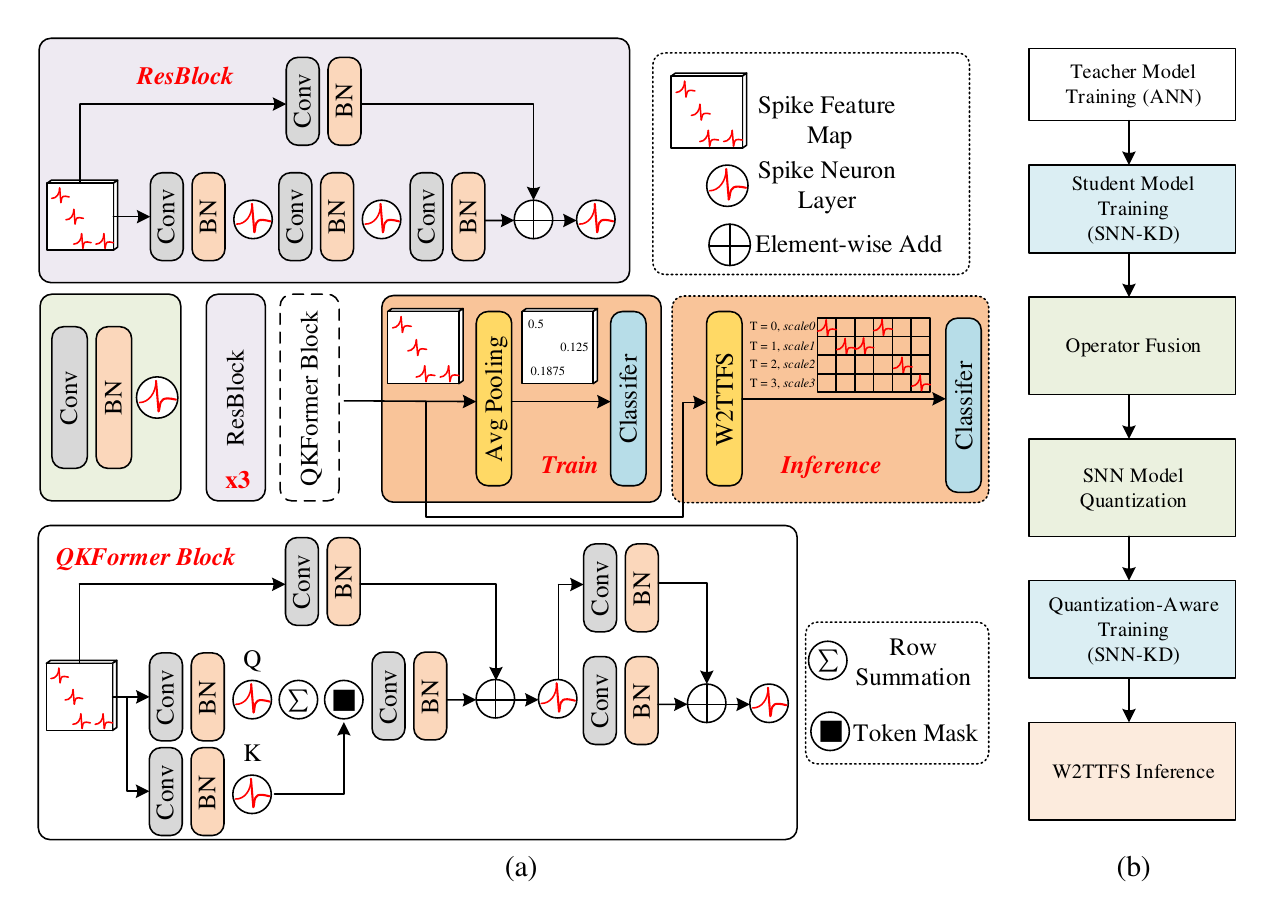} 
\caption{The overview of QKFResNet-11 model and training framework. (a) QKFResNet-11. (b) SNN model training flow based on knowledge distillation.}
\vspace{-14pt}
\label{fig2}
\end{figure}
\vspace{-1ex}\subsection{Window-to-Time-to-First-Spike (W2TTFS)}
As illustrated in Fig. 2(a), to enhance the model performance, we augment the conventional ResNet-11 \cite{chen_high-performance_2023} backbone with QKFormer Blocks\cite{zhou_qkformer_2024} named QKFResNet-11 to incorporate attention mechanisms. The standard average pooling (AP) operation transforms spiking signals into continuous values, leading to non-spiking inputs for the classifier. To tackle this problem, we introduce the W2TTFS method, as depicted in the \textbf{Inference} module in Fig. 2(a), which ensures that the classifier receives spike-based input. The specific conversion flow of W2TTFS is shown in Algorithm 1. Lines 4 and 5 compute the size of the receptive field of the AP operation (denoted as $window\_size$), and initialize a zero matrix with $window\_size^2$ time steps, according to the dimensions of the feature maps produced by the final convolutional layer. Lines 8–16 iterate over the output feature maps by channel and spatial location, identifying the first valid spike timing by counting the number of spikes within each pooling window, as performed in lines 11-13. Finally, lines 17–20 compute the time-dependent weight scaling factors. For example, with $window\_size = 4$, if a valid spike is first emitted at time step $t = 3$, the corresponding scale factor is $3/16$. This factor is subsequently used to scale the weights during the fully connected (FC) computation.
\begin{algorithm}[t]
\caption{Window-to-Time-to-First-Spike (W2TTFS)}
\begin{algorithmic}[1]
\State  \textbf{Given:} spike\_map, which represents output from downsampling convolution layer.
\State  \textbf{Given:} spike\_cnt is the function of getting the number of valid spike.
\State  \textbf{Given:} $T$, $C$, $H_i$, $W_i$, $H_o$, and $W_o$ denote the time steps, channels, input height, input width, output height, and output width, respectively.
\State window\_size $\gets H_i // H_o$
\State spike\_array\_fc $\gets $ \textbf{\textit{Zeros}}($window\_size^2$, $C$, $H_o * W_o$)
\For{$t=0$ \textbf{to} $T$}
    \State spike\_array\_fc.reset()
    \For{$channel = 0$ \textbf{to} $C$}
        \For{$h=0$ \textbf{to} $H_o$}
            \For{$w = 0$ \textbf{to} $W_o$}
                \State \parbox[t]{\dimexpr\linewidth-\algorithmicindent}{
                  \textbf{\textcolor{red}{// window}} \\
                  pooling\_window $\gets$ spike\_map.get($h$,$w$)
                }
                \State vld\_cnt $\gets$ pooling\_window.spike\_cnt()
                \State \parbox[t]{\dimexpr\linewidth-\algorithmicindent}{
                \textbf{\textcolor{red}{// time-to-first-spike}} \\
                spike\_array\_fc[vld\_cnt, channel, h*w] = 1 
                }
             \EndFor
        \EndFor
    \EndFor
    \For{$tt = 0$ \textbf{to} window\_size}
        \State scale = $tt$ / $window\_size^2$
        \State \parbox[t]{\dimexpr\linewidth-\algorithmicindent}{
          spike\_array\_fc[tt].\text{flatten}.classifier(scale)
        }
    \EndFor
\EndFor
\end{algorithmic}
\end{algorithm}
\vspace{-1ex}\subsection{Single-Timestep SNN based on Knowledge Distillation}
In this paper, we introduce a KD-based training framework for SNNs, as illustrated in Fig. 2(b). We first train a high-accuracy ANN as the teacher model. Then, an SNN is constructed as the student model and trained via knowledge distillation. As shown in Fig. 2(a), typical SNN models often include layers such as batch normalization (BN), which pose challenges for hardware deployment. To address this, we apply operator fusion and fixed-point quantization to reduce model complexity and hardware resource usage. Since quantization may degrade accuracy, we further adopt KD-based quantization-aware training (QAT) to mitigate accuracy loss. Finally, during inference, we replace the AP layer with the proposed W2TTFS module to allow full spiking execution.

\vspace{-1ex}\section{NEURAL Architecture}
The NEURAL architecture, as shown in Fig. 3, comprises three key modules: the elastic processing element array (EPA), the pipelined sparse detection array (PipeSDA), and the W2TTFS-based fully connected computing core (WTFC). Weights are streamed into the EPA via elastic W-FIFO, sourced from the weight management unit (WMU). The WMU dynamically schedules the required weights from off-chip memory based on the current computation status and feeds them into the FIFO. Input spikes are handled in a similar manner, when valid spike arrays are buffered in the elastic S-FIFO, the EPA reads them and performs parallel computation with the weights. The PipeSDA module identifies the event receptive field of each input spike based on its coordinates, and maps it to the appropriate sparse detection unit (SDU). The SDU generates a local convolution window, which is forwarded to the EPA for further processing. The WTFC module executes the computation defined by the W2TTFS layer, enabling fully spiking inference at the classifier stage. This end-to-end spiking design allows all computational layers of the SNN model to be efficiently executed within the NEURAL architecture, thus improving system-level integration and execution efficiency.

\begin{figure}[t] 
\centering
\includegraphics[width=\columnwidth]{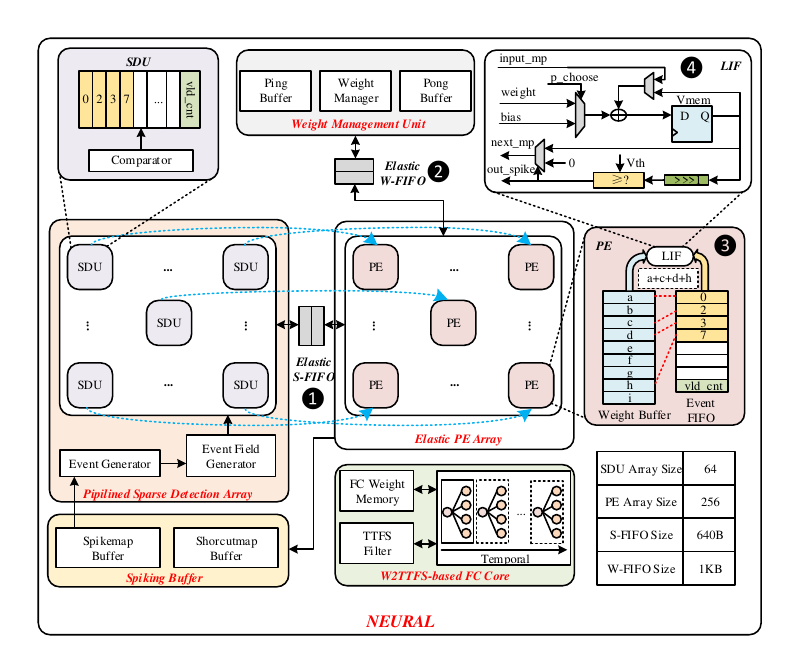} 
\caption{The overall architecture of NEURAL.}
\vspace{-16pt}
\label{fig}
\end{figure}

\vspace{-1ex}\subsection{Hybrid Data-Event Execution Dataflow}
This study proposes a data-event hybrid execution mechanism based on elastic FIFO, which adopts a data-driven control flow at the architectural level while switching to event-driven execution at the granularity of individual neuron computations. \textbf{Data execution.} As illustrated in Fig. 3, the left input of the EPA is the spiking series ❶, and the upper side is the corresponding weight matrix ❷, enabling the system to trigger computation as soon as data from both ends are available, without relying on centralized control. \textbf{Event execution.} As shown in Fig. 3, each PE contains a dedicated event FIFO, the end register of which stores the number of currently valid events, vld\_cnt ❸. During the computation process, the PE reads the event indexes from the FIFO according to the order of vld\_cnt, and obtains the corresponding weight and sends them to the LIF unit, realizing the membrane potential (MP) update computation. The LIF unit updates MP using the corresponding weights and performs threshold comparisons to determine whether to emit a spike ❹. It is fully event-driven, thereby avoiding redundant updates during no-spike intervals.
\vspace{-3ex}\subsection{Pipelined Sparse Detection Array Design}
The main stages of PipeSDA include index generation (IG), center position (CP) generation, and mapping the CP to the SDU array (CP Map) for event detection. \textbf{Index Generation:} as shown in Fig. 4, the system first generates the index of all valid spikes from the input spiking image and stores them in a buffer. \textbf{CP Generation:} CP generation: the CP of the corresponding receptive field for each spike event is calculated based on the generated index. \textbf{CP Map:} these CPs are mapped to specific SDUs in the SDA. Since the coordinates of some CPs may be negative, the virtual SDU is predefined within the SDA architecture to support negative index mapping. Once the mapping is completed, the SDU located at the CP position broadcasts a diffusion signal to its neighboring units (as illustrated on the right of Fig. 4), indicating that the corresponding SDU lies within the active region. Each SDU that receives the signal subsequently updates its internal event FIFO according to the mapping result, preparing for the construction of the convolutional window in later processing.

\begin{figure}[t] 
\centering
\includegraphics[width=\columnwidth]{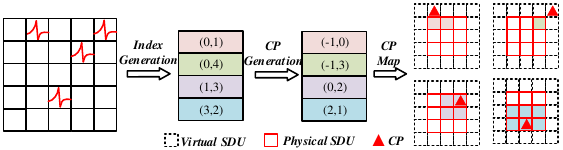} 
\caption{Pipelined sparse detection data flow.}
\label{fig}
\end{figure}

\vspace{-1ex}\subsection{On-the-fly QKFormer Computation}

The proposed NEURAL architecture supports on-the-fly QKFormer computation within the baseline data flow, without requiring a separate spiking transformer unit. As illustrated in Fig. 5, QKFormer operations are directly embedded into the write-back path from the EPA (where spikes are generated by the corresponding PEs) to the Spiking Buffer. As shown in Fig. 5, after computing the Q matrix ❶, we apply bit-wise OR across channels to generate the attention activation state using attention register (atten\_reg) record ❷, corresponding to the Row Summation operation along the Q path in Fig. 2. Subsequently, the K matrix is computed ❸ and written back to the Spiking Buffer. During this process, the atten\_reg is used to determine per-channel activation status (0/1), which is then applied as a QK token mask ❹, aligned with the Token Mask shown in Fig. 2.

\begin{figure}[t] 
\centering
\includegraphics[width=\columnwidth]{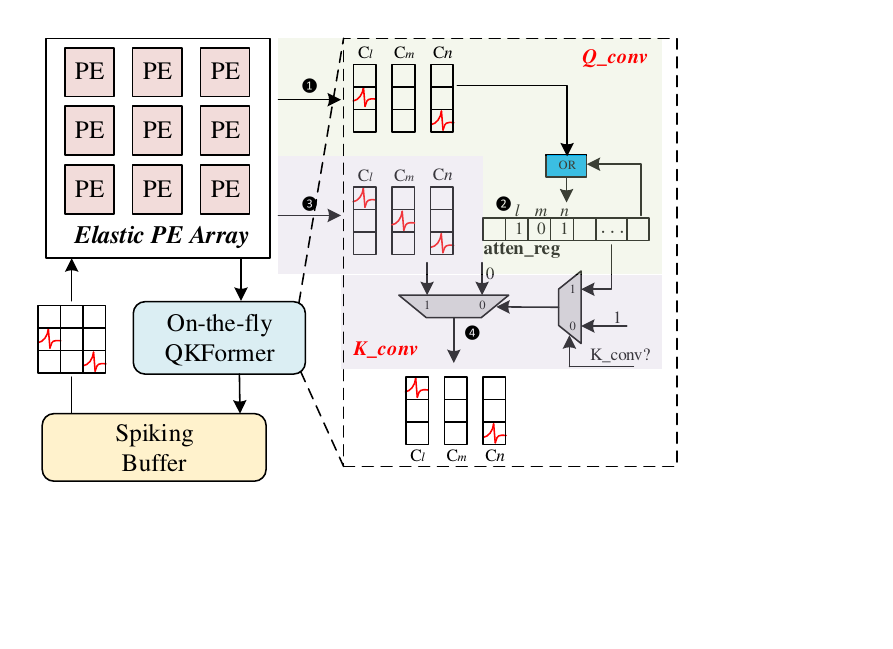} 
\caption{On-the-fly QKFormer computation data flow.}
\vspace{-16pt}
\label{fig}
\end{figure}

\vspace{-3ex}\subsection{W2TTFS-based FC Core}
Fig. 6 shows the  architecture of the W2TTFS-based FC core, which consists of two core modules: the TTFS Filter and the fully-connected computing unit (FCU). The input spiking feature maps are sequentially fed into the TTFS Filter according to the channel order, and the main function of the TTFS Filter is to count the number of valid spikes in each window (denoted as vld\_cnt) according to the size of the pooling window, and generate the corresponding weight scaling factor. However, as shown in Algorithm 1, the scale value may have non-shift friendly decimals (e.g., 3/16). NEURAL performs a fine-grained optimization of the scale generation strategy: the scale no longer depends on the specific spiking location but is uniformly set to the inverse unit of the pooling window (e.g., 1/16). Subsequently, the complex scale is approximated by a time reuse strategy. For example, when the algorithm needs to perform the scaling of 3/16, the system repeats the summation of the unit three times to complete the corresponding membrane potential update, avoiding the multiplication and high-precision division operation. 

\begin{figure}[t] 
\centering
\includegraphics[width=\columnwidth]{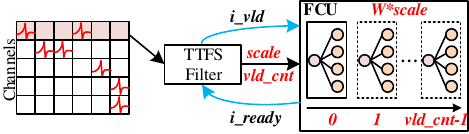} 
\caption{W2TTFS-Based FC core.}
\vspace{-10pt}
\label{fig}
\end{figure}

\section{Experimental Evaluation}
\subsection{Experimental Settings}
\begin{figure}[t] 
\centering
\includegraphics[width=\columnwidth]{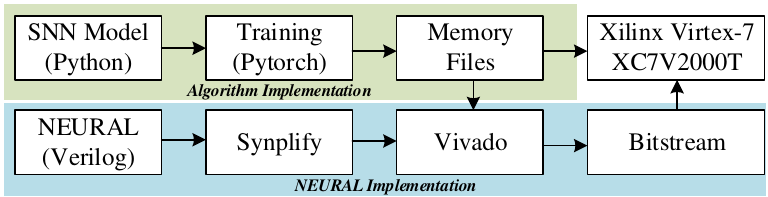} 
\caption{Design flow of algorithm and hardware implementation.}
\vspace{-10pt}
\label{fig}
\end{figure}
\noindent\textbf{Algorithm implementation.} We implemented four SNN models: VGG-11, ResNet-11, QKFResNet-11, and ResNet-19 based on PyTorch and SpikingJelly\cite{fang_spikingjelly_2023}. The LIF neuron decay parameter $\tau$ is 0.5, and the time step is set to 1. The models were trained and tested on CIFAR-10 and CIFAR-100 datasets using an NVIDIA RTX 2080TI. Training is performed using a logit-based KD framework \cite{yu_efficient_2025}, and the teacher model is ResNet-34 with an SGD optimizer with a drive of 0.9, a batch size of 128, and 300 epochs.\\ 
\textbf{Hardware implementation.} After getting a quantized model, the memory files are generated for hardware implementation, as shown in Fig. 7. NEURAL was implemented in Verilog HDL, synthesized using Synplify, and placed and routed using Xilinx Vivado. The design operates at 200MHz on a Xilinx Virtex-7 XC7V2000T FPGA. 

\vspace{-1.5ex}\subsection{Algorithm Analysis}
As illustrated in Fig. 8, it contains four model types: KDT: a full precision model trained using KD; F \& Q: a simplified model applying operator fusion and fixed point quantization; KD-QAT: a QAT model based on KD; and W2TTFS: a hardware-friendly model proposed in this paper. As shown in Fig. 8(a), our KD-trained single-timestep VGG-11 achieves 94.06\% accuracy on CIFAR-10 in the full precision setting, outperforming \cite{chen_sibrain_2024} by 3.01\%, which was evaluated using 4 time steps. After quantization, the KD-QAT VGG-11 exhibits only a 0.17\% accuracy loss, which is 0.34\% smaller than \cite{chen_sibrain_2024}. In CIFAR-100, the improvement is even more significant, reaching 5.77\% and 1.15\%, respectively. In addition, KD-QAT has an obvious advantage in maintaining high accuracy. For example, as shown in Fig. 8(b), the accuracy of ResNet-19 decreases by nearly 7\% after F \& Q, while the loss of accuracy after KD-QAT fine-tuning is only 0.69\%. Specifically, QKFResNet-11 improves accuracy by 0.79\% over ResNet-11 on CIFAR-100 by incorporating the QKFormer Block. Consequently, the KD-based model training and quantization method can effectively improve the accuracy of single-timestep SNN models, reduce the performance loss due to quantization, and thus enhance the robustness of the models.
\begin{figure}[t] 
\centering
\includegraphics[width=\columnwidth]{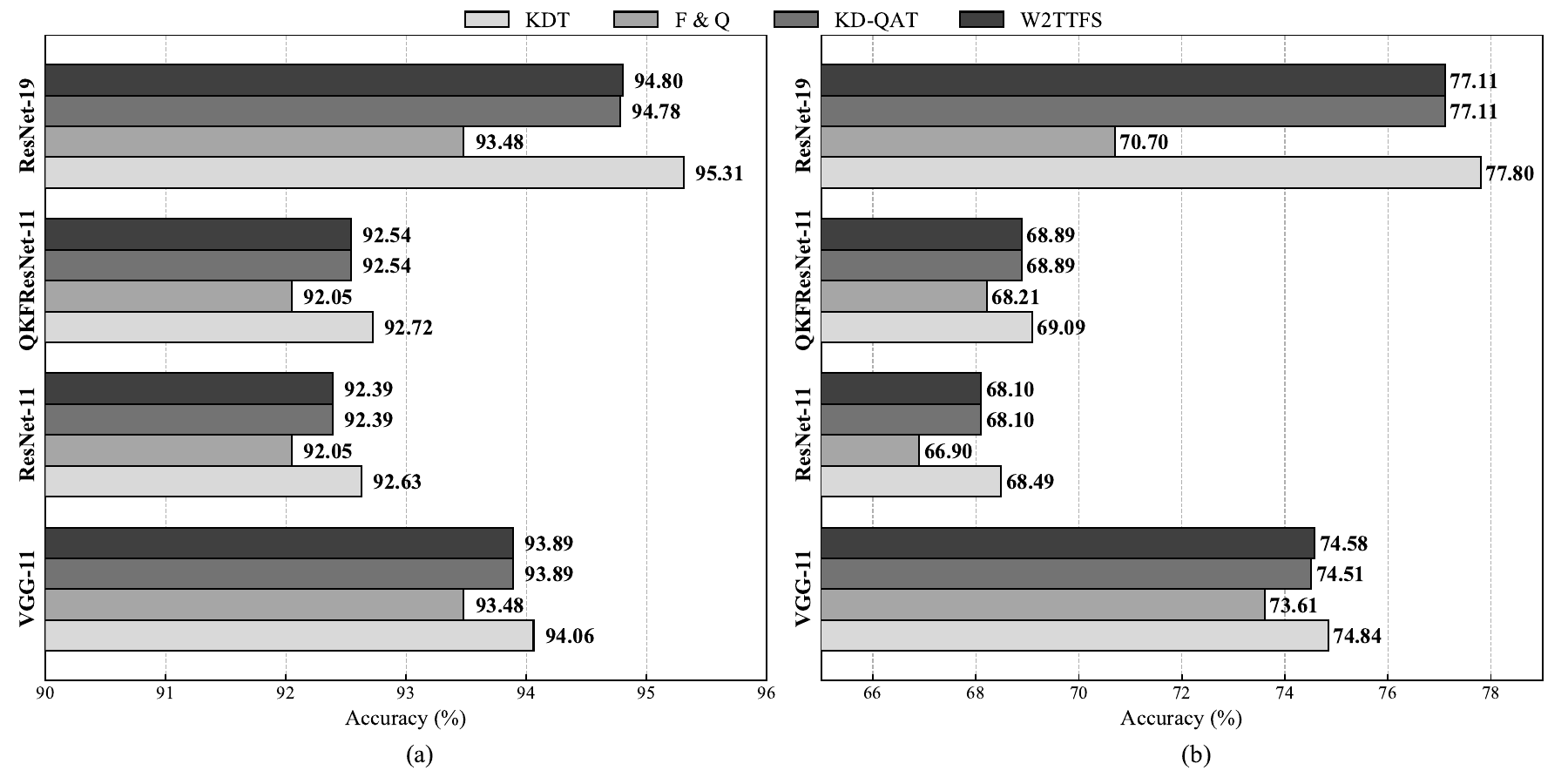} 
\caption{Accuracy of CIFAR-10/CIFAR-100 based on different models. (a) Accuracy of CIFAR-10. (b) Accuracy of CIFAR-100.}
\vspace{-10pt}
\label{fig}
\end{figure}
\begin{table}[t]
  \centering
  \small
  \caption{Hardware Resource Cost of NEURAL}
  \label{tab:resource_breakdown}
  \renewcommand{\arraystretch}{1.2}
  \begin{tabular}{|>{\raggedleft\arraybackslash}p{1.2cm}|>{\raggedleft\arraybackslash}p{1.6cm}|>{\raggedleft\arraybackslash}p{1.5cm}|>{\raggedleft\arraybackslash}p{1.5cm}|>{\raggedleft\arraybackslash}p{1cm}|}
    \hline
    \textbf{Resource} &  \textbf{PipeSDA}&\textbf{EPA}& \textbf{WTFC}& \textbf{Total} \\
    \hline
    LUTs      &  9K (12\%)&33K (45\%)& 1K (1\%)& 74K\\\hline
    Registers &  10K (16\%)&15K (24\%)& 0.7K (1\%)& 63K\\\hline
    BRAM      &  3 (2\%)&64 (47\%)& 25 (18\%)& 137.5\\
    \hline
  \end{tabular}
  \vspace{-6pt}
\end{table}
\vspace{-1ex}\subsection{Computing Resource and Energy Analysis}

\noindent\textbf{Resource Analysis:} 
We deployed the same VGG-11 and ResNet-11 networks used in SiBrain\cite{chen_sibrain_2024} and SCPU\cite{chen_high-performance_2023} on NEURAL and evaluated them on the CIFAR-10 and CIFAR-100 datasets. As shown in Fig. 9, benefiting from the single-timestep execution paradigm, NEURAL consumes only around 70K LUTs, approximately 50\% less than other architectures, and also achieves nearly a 50\% reduction in RAM usage. In terms of recognition accuracy, the VGG-11 model deployed in NEURAL achieves 93.45\% and 72.1\% in CIFAR-10 and CIFAR-100, respectively, which is the highest among all the compared architectures. Consistent results are also observed with the ResNet-11 deployment, as shown in Fig. 9. Table I also reports the resources of each key component. The EPA uses nearly half of the total resources since it is the main computing engine. Sparse processing incurs little hardware overhead because its internal logic is simple. For the proposed WTFC, it uses ~1K LUTs and ~0.7K registers, making it well-suited for edge devices. \\
\textbf{Energy Analysis:} In edge computing, energy consumption directly affects the deployability and life cycle of the system. As shown in Fig. 10, the energy consumption of NEURAL is reduced by nearly 50\% compared to the baseline architecture \cite{chen_sibrain_2024}\cite{chen_high-performance_2023} in the inference process of a single image, and the frame per second (FPS) of image processing is also improved. Specifically, NEURAL achieves a frame rate of 68 FPS while the energy consumption of a single image is only about 10mJ in the VGG-11 (CIFAR-10) task, and the frame rate is increased to 136 FPS in the ResNet-11 (CIFAR-10) task, while the energy consumption remains below 10mJ. The results show that NEURAL is able to achieve a higher energy efficiency ratio while guaranteeing recognition accuracy.

\begin{figure}[t] 
\centering
\includegraphics[width=\columnwidth]{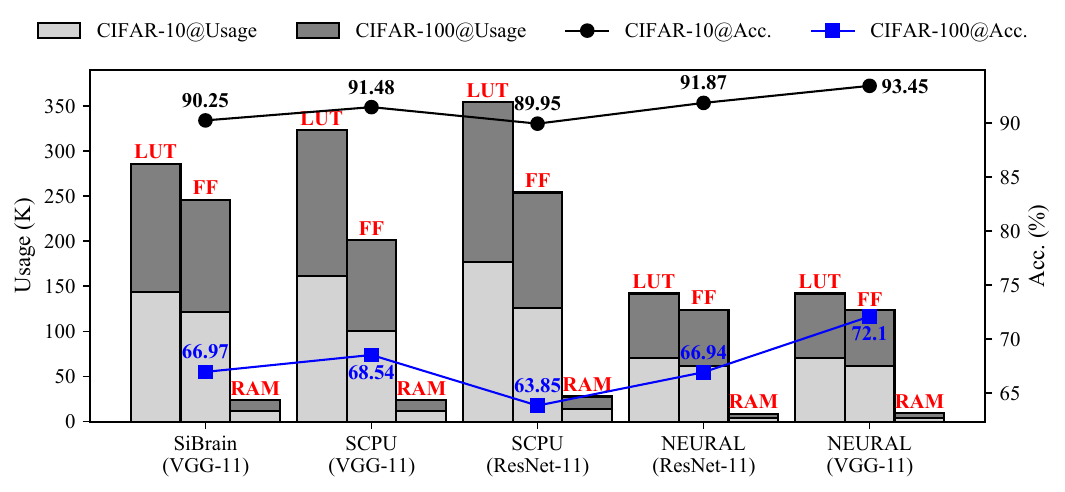} 
\caption{The execution result of VGG-11/ResNet-11 with CIFAR-10/100 on the different platform: resource and accuracy.}
\vspace{-10pt}
\label{fig}
\end{figure}
\begin{figure}[t] 
\centering
\includegraphics[width=\columnwidth]{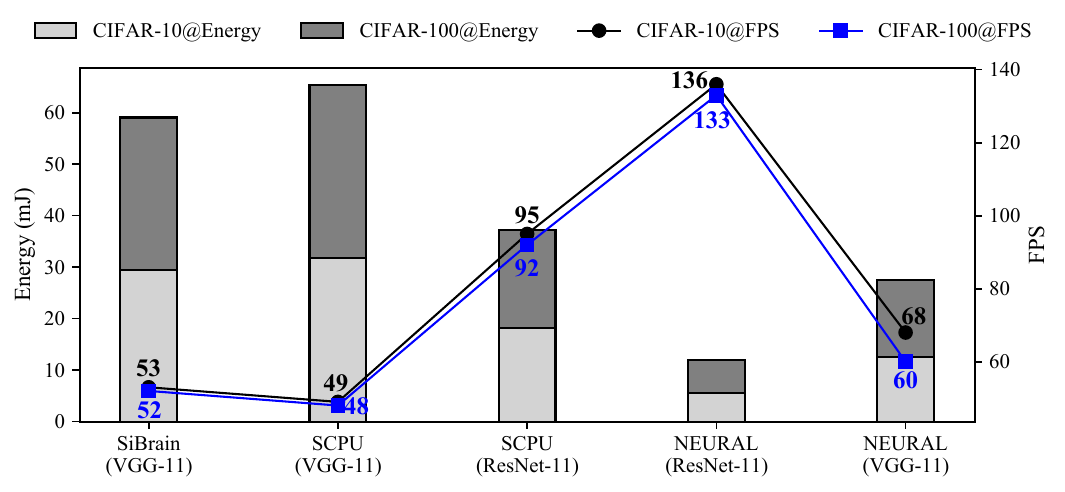} 
\caption{The execution result of VGG-11/ResNet-11 with CIFAR-10/100 on the different platform: computing energy and throught (FPS). }
\vspace{-10pt}
\label{fig}
\end{figure}

\begin{table}[t]
\centering
\caption{ResNet-11 vs. QKFResNet-11 using CIFAR-10/100}
\renewcommand{\arraystretch}{1.5}
\setlength{\tabcolsep}{4pt}
\begin{tabular}{|>{\raggedleft\arraybackslash}p{1.2cm}|>{\raggedleft\arraybackslash}p{1.8cm}|>{\raggedleft\arraybackslash}p{1cm}|>{\raggedleft\arraybackslash}p{1.6cm}|>{\raggedleft\arraybackslash}p{0.85cm}|>{\raggedleft\arraybackslash}p{0.9cm}|}
\hline
\textbf{Data}&\textbf{Model} & \textbf{Total Spikes}& \textbf{Acc. (\%)}& \textbf{Latency (ms)}& \textbf{Energy (mJ)}\\
\hline
\multirow{2}{*}{CIFAR-10}& ResNet-11      & 76K & 91.87           & 7.3 & 5.56 \\
\cline{2-6}
& QKFResNet-11   & 72K & 92.01 (+0.14)& 9.7 & 8.14 \\
\hline
\multirow{2}{*}{CIFAR-100}& ResNet-11      & 83K & 66.94           & 7.5 & 6.44 \\
\cline{2-6}
& QKFResNet-11   & 84K & 68.53 (+1.59)& 9.9 & 8.26 \\
\hline
\end{tabular}
\end{table}

\vspace{-1ex}\subsection{ResNet-11 vs. QKFResNet-11}

Table II depicts the comparison of deploying ResNet-11 and QKFResNet-11 on NEURAL. As shown in Table II, the integration of attention layers in QKFResNet-11 improves classification accuracy across both datasets, with a notable 1.59\% gain on CIFAR-100. Due to the increased network depth, QKFResNet-11 incurs an additional latency of approximately 2 ms. Total Spikes (TS) denotes the total spikes generated during inference. As shown in Table II, QKFormer integration can reduce TS on CIFAR-10 for efficient spike suppression, while increasing TS on CIFAR-100 to adapt to higher task complexity. Importantly, compared to \cite{chen_high-performance_2023}, our QKFResNet-11 achieves a 4.68\% accuracy improvement while reducing energy by 10 mJ.

\vspace{-1.2ex}\subsection{Comparison with Prior Neuromorphic Architectures}
Giga synaptic operations per second per watt (GSOPS/W) is a common metric for evaluating the energy efficiency of SNN hardware architectures. As illustrated in Table III, when deploying ResNet-11 model on the CIFAR-10 dataset using FP8 precision, NEURAL achieves an accuracy of 91.87\%, a frame rate of 136 FPS, a power consumption of just 0.758 W, and an energy efficiency of 46.65 GSOPS/W. Furthermore, with the VGG-11 model on the same dataset, accuracy improves to 93.45\%, with a frame rate of 68 FPS, a power consumption of 0.792 W, and an even higher energy efficiency of 52.37 GSOPS/W. Compared with other state-of-the-art platforms, NEURAL consumes substantially less power than SiBrain \cite{chen_sibrain_2024} (1.56 W) and STI-SNN \cite{wang_sti-snn_2025} (1.34 W), while also achieving superior accuracy and efficiency. Meanwhile, NEURAL achieves up to 3.9× higher computational efficiency than STI-SNN \cite{wang_sti-snn_2025}, even under the same single-timestep setting. For a fair comparison, we adopt the efficiency per kLUTs (GSOPS/W/kLUTs) as the evaluation metric. As shown in Table III, NEURAL achieves the highest normalized efficiency of 0.73. In particular, compared to \cite{chen_cerebron_2022}, NEURAL achieves similar accuracy on CIFAR-10 with only a 0.03\% loss, while significantly increasing the FPS by 46 and improving normalized efficiency by 1.97×.  
\begin{table}[t]
\centering
\caption{Comparison with Existing SNN Accelerators on CIFAR-10} 
\renewcommand{\arraystretch}{1.4}
\setlength{\tabcolsep}{4pt}

\begin{tabular}{|>{\centering\arraybackslash}p{1.4cm}|>{\centering\arraybackslash}p{0.9cm}|c|c|c|>{\centering\arraybackslash}p{1cm}|>{\centering\arraybackslash}p{0.9cm}|}
\hline
\textbf{Platform}  &\textbf{\cite{chen_sibrain_2024}}& \textbf{\cite{chen_cerebron_2022}}& \textbf{\cite{wang_sti-snn_2025}}& \textbf{\cite{aliyev_exploring_2025}} & \multicolumn{2}{|c|}{\textbf{NEURAL}}\\
\hline
\textbf{Device}  &V. 7& Z. 7& Z. U& V. U& \multicolumn{2}{|c|}{V. 7}\\
\hline
\textbf{Fmax (MHz)}  &200 
& 200& 200 & 100 & \multicolumn{2}{|c|}{200}\\
\hline
\textbf{Model}  &VGG-11 
& MobileNet& SCNN5& VGG-9& ResNet-11&VGG-11\\
\hline
\textbf{Precision}  &FP8 
& N/A& INT8 & IN4 & \multicolumn{2}{|c|}{FP8}\\
\hline
\textbf{Acc. (\%)}&90.25 
& 91.90& 90.31 & 86.6 & 91.87&93.45\\
\hline
\textbf{FPS}  &53 
& 90& 397& 120 & 136& 68\\
\hline
\textbf{Power (W)}  &1.56 
& 1.4& 1.53& 0.73 & 0.76&0.79\\
\hline
\textbf{Efficiency} \newline \textbf{(GSOPS/W)}  &84.16 
& 31.6& 13.46& 64.11 & 46.65& 52.37\\
\hline
 \textbf{Norm. Eff}
\textbf{(GSOPS/W/ kLUTs)} &0.60& 0.37& 0.52& 0.58&0.65&0.73\\\hline
\end{tabular}
\vspace{-10pt}
\end{table}

\vspace{-1ex}\section{Conclusion}
In this paper, we proposed NEURAL, a hybrid data-event execution neuromorphic architecture with elastic interconnection, which enabled efficient SNNs execution and supported on-the-fly computation of spiking QKFormer. We introduced a KD-based training framework to achieve competitive accuracy in single time-step execution. For hardware-friendly deployment, the trained SNN model performed operator fusion, quantization, and KD-based QAT. Furthermore, we proposed the W2TTFS mechanism as a replacement for the AP layer, enabling full-spike execution. 
Experimental results demonstrated that NEURAL achieves high accuracy and performance with significantly reduced hardware resource consumption. In the future, we plan to explore more applications on NEURAL, including image segmentation\cite{shaozhen_image_segmantation} and spiking large language models \cite{zhao2025brillm}, to further promote the practical adoption of energy-efficient neuromorphic computing. 

\vspace{12pt}
\balance
\bibliographystyle{ieeetr}
\bibliography{references}

\begin{thebibliography}{10}

\bibitem{fang_energy-efficient_2025}
C.~Fang, Z.~Shen, Z.~Wang, C.~Wang, S.~Zhao, F.~Tian, J.~Yang, and M.~Sawan, ``An energy-efficient unstructured sparsity-aware deep {SNN} accelerator with 3-{D} computation array,'' {\em IEEE Journal of Solid-State Circuits}, vol.~60, pp.~977--989, Mar. 2025.

\bibitem{chen_sibrain_2024}
Y.~Chen, W.~Ye, Y.~Liu, and H.~Zhou, ``{SiBrain}: {A} sparse spatio-temporal parallel neuromorphic architecture for accelerating spiking convolution neural networks with low latency,'' {\em IEEE Transactions on Circuits and Systems I: Regular Papers}, vol.~71, pp.~6482--6494, Dec. 2024.

\bibitem{chen_cerebron_2022}
Q.~Chen, C.~Gao, and Y.~Fu, ``Cerebron: {A} {Reconfigurable} {Architecture} for {Spatiotemporal} {Sparse} {Spiking} {Neural} {Networks},'' {\em IEEE Transactions on Very Large Scale Integration (VLSI) Systems}, vol.~30, pp.~1425--1437, Oct. 2022.

\bibitem{wu_spatio-temporal_2018}
Y.~Wu, L.~Deng, G.~Li, J.~Zhu, and L.~Shi, ``Spatio-temporal backpropagation for training high-performance spiking neural networks,'' {\em Frontiers in Neuroscience}, vol.~12, May 2018.

\bibitem{liang_knowledge_2024}
X.~Liang, G.~Chao, M.~Li, and Y.~Zhao, ``Knowledge distill for spiking neural networks,'' in {\em 2024 {International} {Joint} {Conference} on {Neural} {Networks} ({IJCNN})}, pp.~1--8, June 2024.

\bibitem{yu_efficient_2025}
C.~Yu, X.~Zhao, L.~Liu, S.~Yang, G.~Wang, E.~Li, and A.~Wang, ``Efficient logit-based knowledge distillation of deep spiking neural networks for full-range timestep deployment,'' May 2025.

\bibitem{yao_spike-driven_2023}
M.~Yao, J.~Hu, Z.~Zhou, L.~Yuan, Y.~Tian, B.~Xu, and G.~Li, ``Spike-driven transformer,'' {\em Advances in Neural Information Processing Systems}, vol.~36, pp.~64043--64058, Dec. 2023.

\bibitem{zhou_qkformer_2024}
C.~Zhou, H.~Zhang, Z.~Zhou, L.~Yu, L.~Huang, X.~Fan, L.~Yuan, Z.~Ma, H.~Zhou, and Y.~Tian, ``{QKFormer}: {Hierarchical} spiking transformer using {Q}-{K} attention,'' {\em Advances in Neural Information Processing Systems}, vol.~37, pp.~13074--13098, Dec. 2024.

\bibitem{wang_sti-snn_2025}
K.~Wang, C.~Yang, C.~Yu, Y.~S. Ang, B.~Wang, and A.~Wang, ``{STI}-{SNN}: {A} 0.14 {GOPS}/{W}/{PE} single-timestep inference {FPGA}-based {SNN} accelerator with algorithm and hardware co-design,'' June 2025.

\bibitem{diehl_unsupervised_2015}
P.~U. Diehl and M.~Cook, ``Unsupervised learning of digit recognition using spike-timing-dependent plasticity,'' {\em Frontiers in Computational Neuroscience}, vol.~9, Aug. 2015.

\bibitem{rueckauer_conversion_2017}
B.~Rueckauer, I.-A. Lungu, Y.~Hu, M.~Pfeiffer, and S.-C. Liu, ``Conversion of continuous-valued deep networks to efficient event-driven networks for image classification,'' {\em Frontiers in Neuroscience}, vol.~11, Dec. 2017.

\bibitem{chu_neuromorphic_2022}
H.~Chu, Y.~Yan, L.~Gan, H.~Jia, L.~Qian, Y.~Huan, L.~Zheng, and Z.~Zou, ``A neuromorphic processing system with spike-driven {SNN} processor for wearable {ECG} classification,'' {\em IEEE Transactions on Biomedical Circuits and Systems}, vol.~16, pp.~511--523, Aug. 2022.

\bibitem{nguyen_low-power_2022}
D.-A. Nguyen, X.-T. Tran, K.~N. Dang, and F.~Iacopi, ``A low-power, high-accuracy with fully on-chip ternary weight hardware architecture for deep spiking neural networks,'' {\em Microprocessors and Microsystems}, vol.~90, p.~104458, Apr. 2022.

\bibitem{aliyev_pulse_2024}
I.~Aliyev and T.~Adegbija, ``{PULSE}: {Parametric} hardware units for low-power sparsity-aware convolution engine,'' in {\em 2024 {IEEE} {International} {Symposium} on {Circuits} and {Systems} ({ISCAS})}, pp.~1--5, May 2024.

\bibitem{9773259}
J.-J. Lee, W.~Zhang, and P.~Li, ``Parallel time batching: Systolic-array acceleration of sparse spiking neural computation,'' in {\em 2022 IEEE International Symposium on High-Performance Computer Architecture (HPCA)}, pp.~317--330, 2022.

\bibitem{chen_high-performance_2023}
Y.~Chen, Y.~Liu, W.~Ye, and C.-C. Chang, ``The high-performance design of a general spiking convolution computation unit for supporting neuromorphic hardware acceleration,'' {\em IEEE Transactions on Circuits and Systems II: Express Briefs}, vol.~70, pp.~3634--3638, Sept. 2023.

\bibitem{fang_spikingjelly_2023}
W.~Fang, Y.~Chen, J.~Ding, Z.~Yu, T.~Masquelier, D.~Chen, L.~Huang, H.~Zhou, G.~Li, and Y.~Tian, ``{SpikingJelly}: {An} open-source machine learning infrastructure platform for spike-based intelligence,'' {\em Science Advances}, vol.~9, p.~eadi1480, Oct. 2023.

\bibitem{aliyev_exploring_2025}
I.~Aliyev, J.~Lopez, and T.~Adegbija, ``Exploring the sparsity-quantization interplay on a novel hybrid {SNN} event-driven architecture,'' in {\em 2025 {Design}, {Automation} \& {Test} in {Europe} {Conference} ({DATE})}, pp.~1--7, Mar. 2025.

\bibitem{shaozhen_image_segmantation}
W.~Ye, S.~Chen, H.~Liu, Y.~Liu, Y.~Chen, Y.~Cui, and W.~Lin, ``The architecture design and training optimization of spiking neural network with low-latency and high-performance for classification and segmentation,'' {\em Neural Networks}, vol.~191, p.~107790, 2025.

\bibitem{zhao2025brillm}
H.~Zhao, H.~Wu, D.~Yang, A.~Zou, and J.~Hong, ``Brillm: Brain-inspired large language model,'' {\em arXiv preprint arXiv:2503.11299}, 2025.

\end{thebibliography}

\end{document}